\documentclass{ws-procs975x65}
\input epsf 
\begin{document}

\title{Branons as dark matter
\footnote{\uppercase{C}ontribution to the \uppercase{T}enth 
\uppercase{M}arcel
\uppercase{G}rossmann \uppercase{M}eeting on \uppercase{G}eneral 
\uppercase{R}elativity. \uppercase{T}his work is supported by \uppercase{DGICYT} 
(\uppercase{S}pain) under 
project numbers \uppercase{FPA} 2000-0956 and \uppercase{BFM} 2002-01003}}

\author{J.A.R. Cembranos, A. Dobado and A.L. Maroto}

\address{Departamento de F\'{\i}sica Te\'orica I \\
Universidad Complutense de Madrid, \\ 
28040 Madrid, Spain\\
E-mail:dobado@fis.ucm.es}


\maketitle

\abstracts{
In the brane-world scenario with low tension, brane fluctuations (branons) 
together with the Standard Model particles are the only relevant degrees of 
freedom at low energies. Branons are stable, weakly interacting, massive 
particles and their relic abundance can account for the dark matter of the 
universe. In a certain range of the parameter space, they could be detectable 
by future direct search experiments.}

\section{Brane fluctuations: branons}
In the brane-world scenario our universe is understood as a 3-dimensional 
brane embedded in a  $D$-dimensional space-time ($D = 4 + N$). 
The fundamental scale of gravity in
 $D$ dimensions $M_D$ is no longer the Planck scale $M_P$, but can
be much lower, and the 
tension of the brane $\tau=f^4$ fixes a completely independent
energy scale \cite{ADD}.

In this context, new fields appear on the brane.
On one hand, we have the Kaluza-Klein modes of the gravitons 
propagating in the $D$-dimensional bulk space, and on the other, 
the fields which describe the  brane fluctuations. These fields are called 
branons and they are the Goldstone bosons associated to the spontaneous 
breaking of the  translational invariance in the extra dimensions induced 
by the presence of 
the brane \cite{Sundrum,DoMa}. But, in general, translational 
invariance is not an exact symmetry 
of the bulk space, i.e: 
branons can acquire a mass $M$ \cite{BSky}.
If $f \ll M_D$ (low tension), KK gravitons decouple from the SM particles 
\cite{GB}, and therefore, at low energies, only SM particles 
and branons are important.

The SM-branon low-energy effective Lagrangian was obtained in \cite{ACDM}
and it reads:
\begin{eqnarray}
{\mathcal L}_{Br}&=&
\frac{1}{2}g^{\mu\nu}\partial_{\mu}\pi^\alpha
\partial_{\nu}\pi^\alpha-\frac{1}{2}M^2\pi^\alpha\pi^\alpha\nonumber  \\
&+& \frac{1}{8f^4}(4\partial_{\mu}\pi^\alpha
\partial_{\nu}\pi^\alpha-M^2\pi^\alpha\pi^\alpha g_{\mu\nu})
T^{\mu\nu}_{SM}\label{lag}
\end{eqnarray}

We see that branons interact by pairs with the SM 
energy-momentum tensor. This means that they are stable particles. 
On the other hand, their couplings are suppressed by the 
brane tension $f^4$, i.e. they are weakly interacting. These features
make them natural dark matter candidates \cite{CDM,M} (see \cite{Munoz}
for a recent review on dark matter).

\section{Branons as dark matter}
When the branon annihilation rate, $\Gamma=n_{eq}\langle\sigma_A v
\rangle$,                                 
equals the universe expansion rate $H$, the branon abundance 
freezes out relative to the entropy density. This happens at the 
so called freeze-out temperature $T_f=M/x_f$. We have computed this 
relic branon abundance in two cases: either relativistic branons at 
freeze-out (hot-warm) or non-relativistic (cold), and assuming that 
the evolution of the universe is standard for $T<f$ (see Fig. 1).

\begin{figure}
\centerline{\epsfxsize=12cm\epsfbox{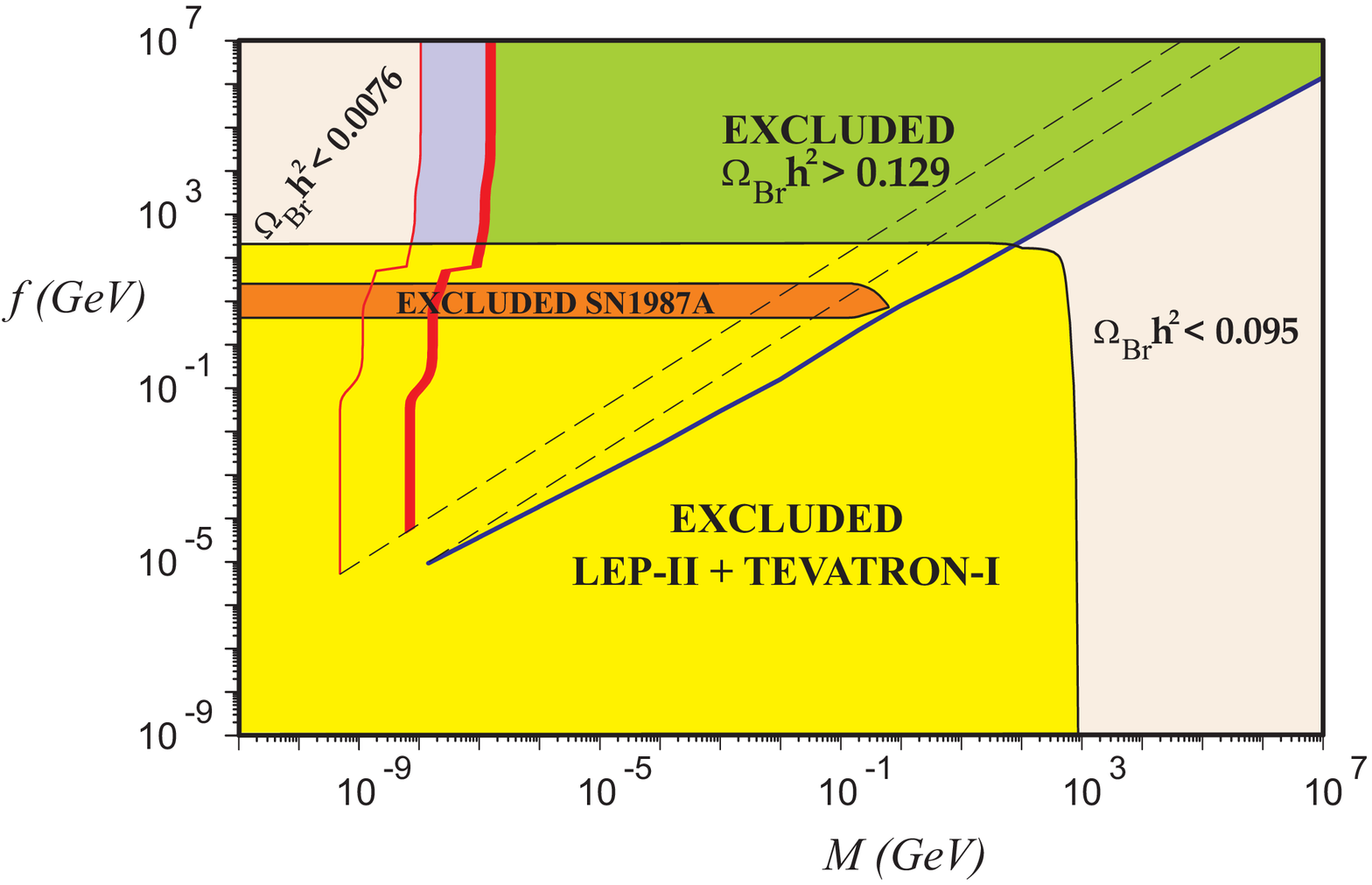}}
\end{figure}

{\footnotesize Relic abundance in the $f-M$ plane for a model with one branon of
mass: $M$. The two lines on the left correspond to the $\Omega_{Br}h^2=0.0076$ and 
$\Omega_{Br}h^2=0.129 - 0.095$ curves for hot-warm relics, whereas the right line
corresponds to the latter limits for cold relics (see \cite{CDMCo} for details). 
The lower area is excluded by single-photon processes at LEP-II \cite{ACDM,L3} together 
with monojet signal at Tevatron-I \cite{baul}. The upper area is also excluded by 
cosmological branon overproduction \cite{CDMCo}. The astrophysical constraints are less
restrictive and they mainly come from supernova cooling by branon emission 
\cite{CDMCo}.}

\section{Detecting branons}

If branons make up the galactic halo, they could be detected by direct search 
experiments from the energy transfer in elastic collisions with nuclei of a 
suitable target. For the allowed parameter region in Fig. 1, branons cannot
be detected by present experiments such as DAMA, ZEPLIN 1 or EDELWEISS.
However, they could be observed by future detectors such as CRESST II, CDMS or
GENIUS.

Branons could also be detected indirectly: their annihilations in the galactic 
halo can give rise to pairs of photons or $e^+ e^-$ which could be detected by  
$\gamma$-ray telescopes such as MAGIC or GLAST or antimatter detectors 
(see \cite{AMS} for an estimation of positron and photon fluxes from
branon annihilation in AMS). 
Annihilation of branons trapped in the center of the sun or the earth can 
give rise to high-energy neutrinos which could be detectable by high-energy 
neutrino telescopes such as AMANDA, IceCube or ANTARES.
These searches complement those in high-energy particle colliders (both
in $e^+ e^-$ and hadron colliders) in which 
real (see Fig. 1) and virtual branon effects could be measured 
\cite{CrSt,ACDM,baul}.

\end{document}